\documentclass[letterpaper, 10pt, conference]{ieeeconf}      
\usepackage{xcolor}
\usepackage{graphicx}
\usepackage{cite}
\usepackage{amsmath}
\usepackage{array}
\usepackage{mathtools}
\usepackage{cuted}
\usepackage{amsfonts}
\usepackage{algorithmic}
\usepackage{graphicx}
\usepackage{textcomp}
\newtheorem{theorem}{Theorem}
\newtheorem{lemma}{Lemma}

\newtheorem{remark}{Remark}
\IEEEoverridecommandlockouts                              

\usepackage{amsmath} 
\usepackage{amssymb}  

\title{\LARGE \bf
Consensus Analysis over Clustered Networks of  Multi-Agent Systems under  External Disturbances
}

\author{Thiem V. Pham$^{1}$, Quynh T. Thanh Nguyen$^1$
\thanks{$^{1}$Thiem V. Pham and Quynh T. Thanh Nguyen are in Thai Nguyen University of Technology (TNUT), Thai Nguyen city, Vietnam . E-mails: \texttt{\{phuthiem,nttquynh$_{-}$dldk\}@tnut.edu.vn}}%
}

\begin{document}

\maketitle
\thispagestyle{empty}
\pagestyle{empty}

\begin{abstract}
This paper studies a consensus  problem of multi-agent systems subjected to  external disturbances over the clustered network. It considers that the agents are divided into several clusters. They are almost all the time isolated one from another, which has a directed spanning tree. The goal of agents achieves a common value. To support interaction between clusters with a minimum exchange of information, we consider that each cluster has an agent, who can exchange information to any agents outside of its cluster at some discrete instants of time. Our main contribution proposes a  consensus protocol, which takes into account the continuous-time communications among agents inside the clusters and discrete-time communication information across clusters. Accordingly, the consensus and the robust $\mathcal{H}_{\infty}$ consensus over the clustered network are respectively analyzed. Thanks to results from matrix theory and algebraic graph theory, we show that the proposed control protocols can solve the problems mentioned above. Finally, a numerical example is given to show the effectiveness of the proposed theoretical results.
\end{abstract}
\section{Introduction}
Analysis and control of the multi-agent system (MASs) have attracted the interest of many researchers in the control community. The MASs is generally referred to as a system composed of a set of dynamical agents that interact through a communication network to reach a coordinated behavior or operation \cite{Nguyen2020, Pham2019b}. Especially, the cooperative control problems of MASs have been extensively investigated in the past two decades because there are many practical applications such as  energy systems, social networks, brain science, epidemic, etc.

The most interesting problem in the cooperative control of MASs is consensus or synchronization. Many verifiable consensus algorithms have been developed based on continuous-time models and discrete-time models in the past decade, where the close relations between the network connectivity and consensus behavior of MASs are constructed, for example in \cite{Jadbabaie2003a, Olfati-Saber2004, Ren2005}. In \cite{Olfati-Saber2004}, the result of the average consensus of continuous-time  MASs with fixed and switching topologies was investigated. It has shown that the strongly connected and balanced directed graphs play a key role in solving average consensus problem. In order to asymptotically convergence analysis of these consensus protocols, the Lyapunov function of disagreement vector is introduced. Moreover, the discrete and continuous consensus algorithms are proposed under dynamically changing topology \cite{Ren2005}. These proved that the consensus can be achieved asymptotically if the union of the directed interaction graphs contains a spanning tree frequently enough.  Note here that, the communication on the aforementioned results is only continuous or discrete.

The question for considering the analysis consensus in a clustered network arises from problems that would benefit from a division of a large network into subnetworks (called clusters) that are almost all the time isolated one from another, such as the energy optimization in the wireless sensor network \cite {Halgamuge2003, Kushner2013}, or the problem of formation of multiple unmanned aerial vehicles \cite{Pham2019c, Pham2020c}, where each cluster is a platoon. In each cluster,  there exists an agent called a leader who can exchange information outside of its cluster at some specific discrete-time, while the interaction among agents inside each cluster happens in continuous-time. 

The main distinctness of this question compared to the analysis consensus in \cite{Jadbabaie2003a, Olfati-Saber2004, Ren2005} lies on two main aspects: 1) The communication among agents in the network either continuous-time or discrete-time, meanwhile the communication in the consideration network is hybrid; 2) The network is composed/ divided of/into by several clusters. This leads to the problem that although each cluster can achieve consensus, the consensus of the overall network is not guaranteed. Thus, some works focus on finding out the sufficient conditions of stability over the clustered network. In particular, the study in \cite{Pham2019a} showed the existence of a sufficient condition to guarantee the overall network asymptotic consensus. In our recent work \cite{Pham2020d,Pham2020c}, a robust formation controller design was proposed for clustered networks of unmanned aerial vehicles. Nevertheless, there is still the loss of mathematical tools to analyze explicitly the consensus over the clustered network under hybrid communication. Moreover, the leader can communicate with any agents outside its cluster instead of with the other leaders. In addition, an external disturbance, and model uncertainty, which are usually considered in the practical networks, are missing on the aforementioned results. 

The first contribution of this paper is to analyze the consensus over the clustered network. These take into account the continuous-time communications among agents within each cluster and discrete update information inter-clusters. Our results are more general than ones in \cite{Pham2020c,Pham2020d,Pham2019a, Bragagnolo2016a}, where the communication across clusters is implemented by only leaders, which expresses in \textit{Remark 1}. Thanks to results from matrix and graph theories, it shows explicitly that the consensus over the clustered network can be solved. Next, a sufficient condition will be derived for the robust $\mathcal{H}_{\infty}$ consensus over the clustered network subjected to external disturbances, which forms the second contribution of this work. 

The rest of the paper is organized as follows. Section II is dedicated to the formulation of the problem under consideration. In Section III, the consensus problem and the robust $\mathcal{H}_{\infty}$ consensus problems of MASs are analyzed. Simulations are given in Section IV. A conclusion sums up this paper.

\textit{Notations and symbols}
The following notations will be use throughout this paper: $\mathbb{N}, \mathbb{R}$, and $\mathbb{R}^+$ stand for the sets of non-negative integers, real, and non-negative real numbers, respectively.  $0_N$ stands for a matrix of zeros with appropriate dimensions. $\textbf{1}_N$ is an $N-$dimensional column vector of 1. We also use $x(t_k^+) = lim_{t\rightarrow t_k, t \geq t_k} x(t)$. 
\section{Preliminaries}
\subsection{Graph Theory}
A directed graph $\mathcal{G}$ with its vertex set $\mathcal{V}=\{v_1,\cdots,v_N\}$ and edge set $\mathcal{E} \subseteq \mathcal{V} \times \mathcal{V}$. Each vertex refers to an agent and each edge $(i,j) \in \mathcal{E}$ corresponds to the communication among the agents $i$ and $j$. The neighboring set of the agent $i$ is represented by $\mathcal{N}_i =\{j\in \mathcal{V}: (i,j) \in \mathcal{E}\}$. Moreover, let $a_{(ij)}$ be the elements of the adjacency matrix $\mathcal A$, defined as $\mathcal A = [a_{(ij)}]$, with $a_{(ij)} >0$ if $ (v_j, v_i) \in \mathcal{E}$, and $a_{(ij)} = 0$ otherwise. The Laplacian matrix $  \mathcal{L}  = [L_{(ij)}]\in \mathbb{R}^{N\times N}$  is defined as $ L_{(ii)}=\sum_{j\neq i} a_{(ij)}; L_{(ij)}=-a_{(ij)}$. The graph is called weighted whenever the elements $a_{(ij)}$ of its adjacency matrix $\mathcal A$ are other than $0-1$. A weighted directed graph has or contains a directed spanning tree if there exists a node called root such that there exists a directed path from this node to every other node. A weighted directed graph is strongly connected if there is a directed path from every node to every other node.

In the following, we consider that the network $\mathcal{G}$ is composed of $m$ weighted directed clusters $\mathcal{C}_{\tau}, \forall \tau \in \{1,\cdots,m\}$ represented by the graphs $\mathcal{G}_1, \cdots, \mathcal{G}_m$ such that $\mathcal{G}_1=(\mathcal{V}_1,\mathcal{E}_1),\cdots,\mathcal{G}_m=(\mathcal{V}_m,\mathcal{E}_m)$, where $\mathcal{V}=\cup_{\tau=1}^m  \mathcal{V}_\tau$ and $V_{\tau} \cap V_g =\varnothing $ for all $\tau,g=1,\cdots m,\tau\neq g$ and $\mathcal{E}=\cup_{\tau=1}^m \mathcal{E}_\tau \cup \mathcal{E}_l$. The graph of each cluster is supposed to be a \textit{directed spanning tree graph}. The communication graph of each cluster $\mathcal{G}_{\tau}$ is represented by a  Laplacian matrix  $\mathcal{L}_{\tau}$. Each cluster  has a specific agent called the leader, and denoted in the following by $l_{\tau } \in \mathcal{V}_{\tau}, \forall \tau \in \{1,\cdots,m\}$. The remaining agents are called followers and are denoted by $f_h$. The set of communication link activated at time $t_k \in \mathbb{N}, t_k \geq 0$ of a time sequence $\{t_k\}$ that satisfies $t_1 <t_2<\cdots, lim_{t_k \rightarrow \infty} t_k =\infty$ with respect to the  leader $l_{\tau}$ is defined as $\mathcal{E}_{l_{\tau}} \neq \varnothing$, where 
\begin{align*}
\mathcal{E}_{l_{\tau}}=\{(l_\tau,h) \in \mathcal{E}| \exists h, l_\tau, h \neq l_\tau, h \in \mathcal V_{l_\tau}\}
\end{align*}
and $\mathcal V_{l_\tau} =\mathcal{V} \cap \mathcal{V}_\tau \cup l_\tau$. A inter-cluster graph $\mathcal{G}_l=(\mathcal{V}_l, \mathcal{E}_l)$, where  $\mathcal{E}_l$ is the set of inter-cluster link activated at time $t_k$ defined as $\mathcal{E}_l=\cup_{\tau=1}^m \mathcal{E}_{l_{\tau}}$ and $\mathcal{V}_l=\cup_{\tau=1}^m \mathcal{V}_{l_\tau}$  is supposed to be  \textit{a strongly connected graph}. 

\subsection{Consensus protocols}
Consider a group of $N$ agents with with integrator dynamics that interact in $m$ clusters. The dynamics of each agent  $i$ is described by
\begin{equation}\label{1}
\dot x_i(t) = u_i(t) + w_i(t),
\end{equation}
where $w_i(t)$ is the external disturbance for agent $i$, which is assumed that $w_i (t)\in \mathfrak{L}_2(0, T]$, where 
\begin{align}\label{27}
	\|w_i(t)\|_T=\left[\int_0^T \|w_i(t)\|^2 dt\right]^{\frac{1}{2}}=\left[\int_0^T w_i^T(t)w_i(t) dt\right]^{\frac{1}{2}}
\end{align}

Our objective now is to design a distributed consensus protocol for a network divided into some clusters, where only a leader of a cluster can interact outside its cluster at some reset times $t_k$. The protocol herein designed considers that  each agent has access to the relative state measurement of its neighbors, and is given by
\begin{align}\label{2}
u_i(t) =\sum_{j=1}^{N} a_{(ij)} \left[x_j(t)-x_i(t)\right],\;\; t \in (t_k,t_{k+1})
\end{align}

Moreover, let us consider that the inter-cluster exchange information at some discrete-time $t_k$ according to 
\begin{align}\label{3}
x_i(t_k^+)=\sum_{j =1}^N P_{e(i,j)}x_j(t_k), \;\;t=t_k
\end{align}
where  $P_e \in \mathbb{R}^{N \times N}$ is a row stochastic matrix associated to the inter-cluster graph, and therefore, the interaction between the leader $l_\tau$ and other agents outside its cluster can be represented as
\begin{align}\label{5}
x_{l_\tau}(t_k^+)=\sum_{j\in \mathcal V_{{l_{\tau}}}} P_{e(l_\tau,j)}x_j(t_k), \;\;t=t_k
\end{align}
where $\mathcal V_{l_\tau} =\mathcal{V} \cap \mathcal{V}_\tau \cup l_\tau$. In addition, according to \eqref{3}--\eqref{5}, a given some $i\neq l_\tau \in \mathcal{C}_\tau$, if $P_{e(i\neq l_\tau,j)}=0$ for all $i \neq j$ then $P_{e(i\neq l_\tau,i)}=1$. It means that there is no jump occuring on the state of the agent $i$ at discrete-time $t_k$.

Then, the collective dynamics of system \eqref{1} under the consensus protocol \eqref{2} and the interaction the inter-cluster \eqref{3} can be rewritten  as

\begin{align}\label{7}
\left\{
\begin{aligned}
&\dot x_i(t)=-\sum_{j=1}^{N} L_{(ij)} x_j(t) + w_i(t),\;\; t \in (t_k,t_{k+1})\\
&x_i(t_k^+)=\sum_{j =1}^N P_{e(i,j)}x_j(t_k), \;\;t=t_k
\end{aligned}
\right.
\end{align}
where $L_{(ij)}$ are element of the Laplacian matrix $\mathcal{L}$. 
\subsection{Some useful lemmas}
In the sequel, the following \textit{Lemmas} are recalled.
\begin{lemma}\label{l3}\cite{Ren2005}
	Let $\Gamma$ be a compact set consisting of $n \times n$ SIA matrices with the property that for any non-negative integer $k$ and any $B_1,\cdots,B_k \in \Gamma$, the matrix product $\prod_{i=1}^k B_i$ is SIA. Then, for given any infinite sequence $B_1,B_2,\cdots$, there exits a column vector $c^\mathsf T$ such that $\lim_{k \rightarrow \infty}\prod_{i=1}^k B_i=\mathbf{1}c^\mathsf T$.
\end{lemma}
\begin{lemma}\label{l4}\cite{Jadbabaie2003a}
	If $B=[b_{ij}]_{n \times n}$ is a stochastic matrix with positive diagonal elements, and the graph associated with $B$ has a spanning tree, then $B$ is SIA.
\end{lemma}
\begin{lemma}\label{l5}\cite{Ren2005}
	For any $t >0$, $e^{-\mathcal{L}t}$ is a stochastic matrix with positive diagonal entries, where $\mathcal{L}$ is the Laplacian of graph $\mathcal{G}$.
\end{lemma}
\begin{lemma}\cite{Jadbabaie2003a}\label{l6}
	Let $m \geq 2$ be a positive integer and let $D_1,\cdots,D_m$ be non-negative $n \times n$ matrices with positive diagonal elements, then 
	\begin{align*}
	D_1D_2 \cdots D_m \geq \gamma (D_1 +D_2+\cdots+D_m),
	\end{align*}
	where $\gamma > 0$ can be specified from $D_i, i=1,\cdots,m$
\end{lemma}
\section{System Analysis}
\subsection{Consensus Problem Analysis with $w_i(t)=0$}
In this subsection, consensus problems in the clustered network of MASs are considered. Now we are in the position to present one of the main results.
\begin{theorem}
	Consider the collective dynamic system \eqref{7} with $w_i(t) = 0$ and assume that impulsive intervals ($t_k, \;t_{k+1})$ for $k \in \mathbb N$  are uniformly bounded, that is, there exist positive constants $\delta_{min}$ and $\delta_{max}$ such that $\delta_{min} \leq  t_{k+1} -t_k \leq  \delta_{max}$ for all $k \in \mathbb N$. Then,  the proposed control protocol \eqref{2} guarantees that the collective dynamics system \eqref{7} achieves consensus.
\end{theorem}
\begin{proof}
Let $x=[x_1, x_2,\cdots,x_N]^\mathsf T \in \mathbb{R}^{N}$, then the system \eqref{7} can be rewritten in a compact form
	\begin{align}\label{10}
	\left\{
    \begin{aligned}
	&\dot x(t)=-\mathcal L x(t),\;\;\;\;\; t \in (t_k,t_{k+1})\\
	&x(t_k^+)=P_e x(t_k), \;\;t=t_k,
	\end{aligned}
	\right.
	\end{align}
where $\mathcal{L} \in \mathbb{R}^{N \times N}$ is the Laplacian matrix associated with the graph $\mathcal{G}$ 
	\begin{equation}\label{11}
	\mathcal{L}=Diag\{\mathcal{L}_1,\cdots,\mathcal{L}_m\}
	\end{equation}
	Then, for any initial condition $x(t_0)=x_0$, the solution of \eqref{10} is obtained by
	\begin{align}\label{12}
	x(t)=e^{-\mathcal L (t-t_k)}x(t_k^+),
	\end{align}
	and $x(t_k^+)$ can be expressed as
	\begin{align}\label{13}
	x(t_k^+)&=\lim_{k \rightarrow \infty} \prod_{i=1}^k P_e e^{-\mathcal L (t_i -t_{i-1})}x(t_0).
	\end{align}
	Moreover,  one has
	\begin{align}
	\sum_{j=1}^{N}P_{e(ij)}=1,
	P_{e(ij)} >0 \label{14}.
	\end{align}
	Subsequently, by employing \eqref{14}, we see that the matrix $P_e \in \mathbb{R}^{N \times N}$ is a row stochastic matrix with positive diagonal elements.
	
	Next, according to \textit{Lemma \ref{l5}}, $e^{-\mathcal L t}$ is a stochastic matrix with positive diagonal entries. Moreover, it follows from \textit{Lemma \ref{l6}} that $P_e e^{-\mathcal L t} \geq \gamma (P_e+e^{-\mathcal L t})$, where $\gamma$ is a positive constant. Therefore, the matrix $P_e e^{-\mathcal L t}$ is a positive diagonal entries.
	
	Due to the inter-cluster communication graph  $\mathcal{G}_l=(\mathcal{V}_l, \mathcal{E}_l)$ is supposed to a strongly connected directed graph. Thus, according to \eqref{14}, we can deduce that the graph of $P_e$ has at least  spanning tree. Moreover, since $P_e e^{-\mathcal L t} \geq \gamma (P_e+e^{-\mathcal L t})$, and the graph of $e^{-\mathcal L t}$ have at least one spanning tree, one can remark that $P_e e^{-\mathcal L t}$ has a spanning tree.
	
	Based on the above analysis, we showed that the matrix $P_e e^{-\mathcal L t}$ is a row stochastic matrix with positive diagonal elements and that its graph has a spanning tree. Then, according to \textit{Lemma \ref{l4}}, the  matrix $P_e e^{-\mathcal L t}$ is SIA.
	
	Therefore, from \textit{Lemma \ref{l3}}, there exits a column vector $c^\mathsf T$ such that
	\begin{align}\label{15}
	\lim_{k \rightarrow \infty} \prod_{i=1}^k P_e e^{-\mathcal L (t_i -t_{i-1})} =\mathbf{1}_N c^\mathsf T.
	\end{align}
	Moreover, $e^{-\mathcal L (t-t_k)}$ is a stochastic matrix, which means that $e^{-\mathcal L (t-t_k)}\mathbf{1}_N =\mathbf{1}_N$. 
	Then, according to \eqref{15}, one has
	\begin{align}\label{16}
	&e^{-\mathcal L (t-t_k)}\left[	\lim_{k \rightarrow \infty}\prod_{i=1}^k P_e e^{-\mathcal L (t_i -t_{i-1})}-\mathbf{1}_N c^\mathsf T\right]=0\nonumber\\
	\Leftrightarrow &e^{-\mathcal L (t-t_k)}	\lim_{k \rightarrow \infty}\prod_{i=1}^k P_e e^{-\mathcal L (t_i -t_{i-1})}-\mathbf{1}_N c^\mathsf T =0,
	\end{align}
	it follows that 
	\begin{align} \label{17}
	e^{-\mathcal L (t-t_k)}	\lim_{k \rightarrow \infty}\prod_{i=1}^k P_e e^{-\mathcal L (t_i -t_{i-1})}\rightarrow \mathbf{1}_N c^\mathsf T.
	\end{align}
	Finally, according to (\ref{17}), \eqref{12} and \eqref{13}, one has
	\begin{align}\label{18}
	\lim_{t \rightarrow \infty}(x(t)-\mathbf{1}_Nc^\mathsf Tx_0) \rightarrow 0,
	\end{align}
	or 
	\begin{align}\label{19}
	\lim_{t \rightarrow \infty}(x_i(t)-\underbrace{c^\mathsf Tx_0}_{x^{\star}}) \rightarrow 0,
	\end{align}
	which implies that the system  \eqref{7} can achieve the consensus. This completes the proof.
\end{proof}
\begin{remark}
    In the particular case, the inter-cluster interaction is done by only leaders, who  interact at some discrete-time $t_k$ through predefined graph $\mathcal{G}_l=(\mathcal{I}, \mathcal{E}_l)$, where $\mathcal{I}=\{l_1,\cdots,l_m \}$ and $\mathcal{E}_l \subset \mathcal{I} \times \mathcal{I}$ (more information please see in \cite{Pham2019c}) such as
	\begin{equation}\label{20}
	x_{l_\tau}(t_k^+)=\sum_{j=1}^{m}P_{l(l_\tau,j)}x_{l_j}(t_k), t=t_k
	\end{equation}
	where $P_l \in \mathbb{R}^{m \times m}$ is a row stochastic matrix associated to the graph $\mathcal{G}_l$ 
	\begin{equation}\label{21}
	\left\{ 
	\begin{aligned}
	&P_{l(l_\tau,j)}=0,  \;\;\;\; if \;(l_\tau,j) \notin \mathcal{E}_l \\
	&P_{l(l_\tau,j)}>0,  \;\;\;\; if \;(l_\tau,j) \in \mathcal{E}_l;i\neq j \\
	&\sum_{j=1}^{m}P_{l(l_\tau,j)}=1,  \;\;\;\; \forall l_\tau=1,\cdots,m\\
	\end{aligned}
	\right. 
	\end{equation}
Then, we can represent the interaction of the inter-cluster by using a extended stochastic matrix $P_e$ as follows
	\begin{align}\label{22}
	P_e=\mathcal{M}^\mathsf T\left[ {\begin{array}{*{20}{c}}
		{{P_l}}&{{0}}\\
		{{0}}&{{I_{N-m}}}\\
		\end{array}} \right]\mathcal{M} \in \mathbb{R}^{N \times N},
	\end{align}
	where $\mathcal{M}$ is a permutation matrix. Thus, the equation \eqref{20} can be expressed by
	\begin{equation}
	x_{i}(t_k^+)=\sum_{j=1}^{N}P_{e(ij)}x_{j}(t_k), \;\;t=t_k,
	\end{equation}
Therefore, we achieve a consensus in the clustered network by using the same analysis above. Moreover, according to \eqref{15} and \eqref{19}, one sees that the column vector $c^\mathsf T$ depends on the communication between clusters $P_e$ and the graph of each cluster $\mathcal{L}$. Therefore,  the final consensus value of the clustered network (see  in \cite{Pham2019c} with the case of simple integrator $\dot{x}_i(t) =u_i(t)$) is 
	\begin{align}\label{23}
	x^{\star}=\underbrace{\frac{\phi^\mathsf T \mathcal{Q}}{\sum_{\tau=1}^{m}\phi_{\tau}}}_{c^\mathsf T}x_0, 
	\end{align}  	
\end{remark}
\subsection{Consensus problem with external disturbances}
In the following, we recall the collective dynamical system under external disturbaces, which is discribed as
\begin{align}\label{28}
\left\{
\begin{aligned}
&\dot x_i(t)=-\sum_{j=1}^{N} L_{(ij)} x_j (t)+w_i(t),\; t \in (t_k,t_{k+1})\\
&x_i(t_k^+)=\sum_{j =1}^N P_{e(i,j)}x_j(t_k), \;\;t=t_k
\end{aligned}
\right.
\end{align}
where $L_{(ij)}$ are element of the Laplacian matrix $\mathcal{L}$.

According to the results in Section III.A, let's introduce  
\begin{align*}
	\psi_i(t)=x_i(t)-c^\mathsf{T}x_0,
\end{align*} 
where $c^\mathsf{T}$ is given in \eqref{18}, and note that $\psi_i(t_k^+)=x_i(t_k^+)-c^\mathsf{T}x_0$. Then, the system \eqref{28} can be rewritten as
\begin{align}\label{29}
\left\{
\begin{aligned}
&\dot \psi_i(t)=-\sum_{j=1}^{N} L_{(ij)} \psi_j (t)+w_i(t),\; t \in (t_k,t_{k+1})\\
&\psi_{i}(t_k^+)=\sum_{j=1}^{N}P_{e(ij)}\psi_{j}(t_k), \;\;t=t_k,
\end{aligned}
\right.
\end{align}
\begin{remark}
	According to $\psi_i(t)=x_i(t)-c^\mathsf{T}x_0$, it is not difficult to recognize  that $\psi_i =0$ if and only if $x_1 =\cdots =x_N=c^\mathsf{T}x_0$. Therefore, we need only to prove that the system (\ref{29}) is asymptotically stable with the disturbance $w(t)$.
\end{remark}

In the following, to suppress the external disturbances in the clustered network, we define an output function $z_i(t)=\psi_i(t)$, then  the system \eqref{29} can be described as
\begin{align}\label{30}
\left\{
\begin{aligned}
&\dot \psi_i(t)=-\sum_{j=1}^{N} L_{(ij)} \psi_j(t) +w_i(t),\; t \in (t_k,t_{k+1})\\
&\psi_{i}(t_k^+)=\sum_{j=1}^{N}P_{e(ij)}\psi_{j}(t_k), \;\;t=t_k,\\
&z_i(t)=\psi_i(t), \;\psi_i(t_0^+)=\psi_{i0}
\end{aligned}
\right.
\end{align}
where $z_i(t)$ is the controlled output and $t_0$ is the initial time, and let $\psi=[\psi_1,\psi_2,\cdots,\psi_N]^\mathsf T \in \mathbb{R}^N$, then the system \eqref{30} can be rewritten in a compact form
\begin{align}\label{31}
\left\{
\begin{aligned}
&\dot \psi(t)=-\mathcal{L}\psi(t) +w(t),\; t \in (t_k,t_{k+1})\\
&\psi(t_k^+)=P_{e}\psi(t_k), \;\;t=t_k,\\
&z(t)=\psi(t), \;\psi(t_0^+)=\psi_0
\end{aligned}
\right.
\end{align}
Finally, the robust $\mathcal{H}_{\infty}$ problem (\textit{the controlled output $z(t)$ satisfies $\|z(t)\|_T \leq \rho \|w(t)\|_T$ with $\rho >0$ for any nonzero $w(t) \in \mathfrak{L}_2(0, T]$})  is equivalent to the performance index
\begin{align}\label{39k}
J=\int_{0}^{T} \left(z(t)^\mathsf Tz(t) -\rho^2 w(t)^\mathsf Tw(t) \right) dt <0,
\end{align}
which can be dealt with in the following theorem.
\begin{theorem}
	Consider the overall collective dynamical system \eqref{31}. If there exits positive scalars $\alpha >0, \rho >0, 0 <\beta < 1$ and positive-definite matrices $P$ such that 
	\begin{align}
	\begin{bmatrix}\label{31t}
    -\mathsf P  \mathcal{L} -\mathcal{L}^\mathsf T \mathsf P  +\alpha \mathsf P &\mathsf P^\mathsf T\\
    \mathsf P&-\rho^2
	\end{bmatrix}<0\\
	\begin{bmatrix}\label{32}
    - \beta \mathsf P & P_{e}^\mathsf T\\
    P_{e} & -\mathsf P^{-1}
	\end{bmatrix}<0
	\end{align}
	then 
	\begin{itemize}
		\item [1)] The controlled output $z(t)$ satisfies  $J< 0$ for zero-initial condition and for all nonzero $w_i(t) \in \mathfrak{L}_2(0, T]$. 
		\item[2)] When $w(t)=0$ the system \eqref{31} is stable with exponential convergence rate $\eta=\alpha -\frac{ln \beta}{T}$.
	\end{itemize}
\end{theorem}
\begin{proof}
	To prove the stability of the (\ref{31}), let's consider the  candidate Lyapunov function
	\begin{align}\label{32t}
	V(t)=V(\psi(t))=\psi^\mathsf T(t) \mathsf P \psi(t)
	\end{align}
 It is sufficient to prove that  the Lyapunov function is satisfied:
	\begin{itemize}
		\item [i)] At some discrete-time $t_k$, one has
	\begin{align}\label{33t}
	V(t_k^+)\leq \beta V(t_k) \;\text{or} \; \lim_{t \rightarrow t_k^+} V(t) \leq \beta V(t_k)
	\end{align}
	where $0 <\beta < 1$. 
	\item[ii)] Between impulses $t_k$ and $t_{k+1}$, for a prescribed scalar $\rho > 0$, the Lyapunov function satisfying
	\begin{align}\label{34t}
	\dot{V}(t)+\alpha V(t)+z(t)^\mathsf Tz(t) -\rho^2 w(t)^\mathsf Tw(t) <0. 
	\end{align}
\end{itemize}

In the sequel, we will show the conditions \eqref{33t} and  \eqref{34t} are equivalent to the performance index \eqref{39k} by giving $T>0$. For any given $T \in (t_k, t_{k+1}]$, 	it follows from \eqref{34t} that 
\begin{align}\label{35t}
\int_{t_k}^{t_{k+1}}\left(\dot{V}(t)+\alpha V(t)+z^\mathsf T(t)z(t)-\rho^2 w^\mathsf T(t)w(t \right) dt <0, 
\end{align}
and using \eqref{35t} successively  on each interval from $t_0$ to $T$ with $\psi(t_0)=0$, one obtains
	\begin{align}\label{36t}
	\begin{aligned}
	&\sum_{v=1}^{k} \int_{t_{v-1}}^{t_{v}} \dot{V}(t)dt+\int_{t_{k}}^{T}\dot{V}(t)dt+
	\sum_{v=1}^{k} \int_{t_{v-1}}^{t_{v}} \alpha V(t)dt\\
	&+\int_{t_{k}}^{T}\alpha V(t)dt+\int_{t_{0}}^{T} \left(z^\mathsf T(t)z(t) -\rho^2 w^\mathsf T(t)w(t) \right) dt <0
	\end{aligned}
	\end{align}
Moreover, it follows from \eqref{33t} and $0 <\beta < 1$, one has $V(t_k^+)- V(t_k)\leq -(1-\beta)V(t_k) <0$ and yields
	\begin{align}\label{37t}
		\begin{aligned}
	&\int_{t_{0}}^{t_{1}} \dot{V}(t)dt+\cdots+\int_{t_{k-1}}^{t_{k}} \dot{V}(t)dt+\int_{t_{k}}^{T} \dot{V}(t)dt\\
	&=V(T)-\sum_{v=0}^{k}\left(V(t_v^+) -V(t_v)\right) >0.
		\end{aligned}
	\end{align}
	Therefore, 
	\begin{align}\label{38t}
		\begin{aligned}
	&\sum_{v=1}^{k} \int_{t_{v-1}}^{t_{v}} \dot{V}(t)dt+\int_{t_{k}}^{T}\dot{V}(t)dt+\\
	&+\sum_{v=1}^{k} \int_{t_{v-1}}^{t_{v}} \alpha V(t)dt +\int_{t_{k}}^{T}\alpha V(t)dt>0.
\end{aligned}
	\end{align}
	It follows from \eqref{36t} and \eqref{38t} that 
	\begin{align}\label{39t}
	\int_{t_{0}}^{T} \left(z^\mathsf T(t)z(t) -\rho^2 w^\mathsf T(t)w(t) \right) dt <0.
	\end{align}
	
	In the following, with $t \in (t_k, \;t_{k+1})$ the derivative of Lyapunov function with respect to \eqref{31} is 
	\begin{align}\label{40t}
	\begin{aligned}
	\dot{V}(t)= &\psi^\mathsf T(t) (-\mathsf P \mathcal{L} -\mathcal{L}^\mathsf T \mathsf P ) \psi(t) \\
	&+ \psi^\mathsf T(t) \mathsf P  w(t) + \psi(t) \mathsf P  w^\mathsf T(t)
	\end{aligned}
	\end{align}
	and by using \eqref{34t}, one has
	\begin{align}\label{41t}
	&\psi^\mathsf T (-\mathsf P  \mathcal{L} -\mathcal{L}^\mathsf T \mathsf P  +\alpha \mathsf P ) \psi +\nonumber\\
	&+\psi^\mathsf T \mathsf P  w + \psi \mathsf P  w^\mathsf T +\psi^\mathsf T \psi -\rho^2 w^\mathsf Tw <0.\\
	\Leftrightarrow &-\left[\rho w -\frac{1}{\rho} \mathsf P  \psi\right]^\mathsf T\left[\rho w -\frac{1}{\rho} \mathsf P \psi \right]+\nonumber\\
	&+\psi^\mathsf T (\frac{1}{\rho^2} \mathsf P ^\mathsf T \mathsf P  -\mathsf P  \mathcal{L} -\mathcal{L}^\mathsf T \mathsf P  +\alpha \mathsf P ) \psi <0.
	\end{align}
	In order to ensure the condition \eqref{34t}, one has
	\begin{align}\label{42t}
	\frac{1}{\rho^2} P^\mathsf T \mathsf P  -\mathsf P  \mathcal{L} -\mathcal{L}^\mathsf T \mathsf P  +\alpha \mathsf P  <0
	\end{align}
	and \eqref{31t} is obtained by using Shur complement .
	
	On the other hand, at the reset time $t=t_k$, one has
	\begin{align}\label{43t}
	V(t_{k}^+)-\beta V(t_k)=\psi^\mathsf T(t_k)(P_{e}^\mathsf T\mathsf P P_{e}- \beta \mathsf P )\psi(t_k)
	\end{align}
	Then, to guarantee the condition \eqref{32t}, one needs
	\begin{align}\label{44t}
	P_{e}^\mathsf T\mathsf P P_{e}- \beta \mathsf P <0.
	\end{align}
	and  \eqref{32} is obtained by using again Shur complement.This completes the proof of part 1.
	 
	In case of $w(t)=0$, from \eqref{34t}, one obtains
	\begin{align}\label{45t}
	\dot{V}(t)+\alpha V(t) <0, \; t\in (t_k, t_{k+1})
	\end{align}
	which obtains that
	\begin{align}\label{46t}
	{V}(t)<V(t_k^+) e^{- \alpha (t-t_{k+1})} , \; t\in (t_k, t_{k+1})
	\end{align}
	and at reset time $t=t_k^+$,
	\begin{align}\label{47t}
	V(t_k^+) \leq \beta V(t_k)
	\end{align}
	According to the results from \eqref{46t} and \eqref{47t}, one has 
	\begin{align}\label{48t}
	V(t_1) \leq V(t_0^+)e^{- \alpha (t_1-t_0)} , \; t\in (t_0, t_{1})
	\end{align}
	and 
	\begin{align}\label{49t}
	V(t_1^+) \leq \beta V(t_1) \leq V(t_0^+)\beta e^{-\alpha (t_1-t_0)} , \; t=t_1^+
	\end{align}
	In general, for $t \in (t_k, t_{k+1})$, 
	\begin{align}\label{52t}
	V(t_k^+)<V(t_0^+) \beta ^k e^{- \alpha (t_{k+1}-t_0)}, 
	\end{align}
	and using \eqref{47t}, one has
	\begin{align}\label{53t}
	V(t)<V(t_0^+) \beta^k e^{- \alpha (t-t_0)} ,
	\end{align}
	Denoting $T$ as the average impulsive interval of the impulsive sequence $\{t_1,t_2,...\}$, one has $k\geq \frac{t-t_0}{T}-N_0$, where $N_0$ is a positive integer \cite{Lu2010}, then
	\begin{align}\label{54t}
	V(t)&<V(t_0^+) \beta^{\frac{t-t_0}{T}-N_0} e^{- \alpha (t-t_0)} ,\nonumber\\
	&<V(t_0^+) \beta^{-N_0}e^{- (\alpha -\frac{ln \beta}{T})(t-t_0)} 
	\end{align}
	It means that the system \eqref{31} is stable with $w(t)=0$. This completes the proof of part 2. 
\end{proof}
\begin{remark}
    $-\mathcal{L}$ has a $m$ unique zero eigenvalues and other $N-m$ eigenvalues which has negative a real parts \cite{Pham2019a}, then there always exist the solution for LMIs \eqref{31t}.
\end{remark}
\section{Simulation Results}
In this section, numerical simulation is given to validate the results reported in the previous section. The network contains 7 agents and it is partitioned into 2 clusters having 4 and 3 elements, respectively. Moreover, let us suppose that the graphs of clusters $\mathcal{C}_1$ and $\mathcal{C}_2$  are respectively represented by the Laplacian matrices $\mathcal{L}_1$ and  $\mathcal{L}_2$
\begin{align*}
\mathcal{L}_1=
\begin{bmatrix}
   2 &-1&0&-1\\
-1 &2&-1&0\\
-1 &-1&2&0\\
-1&-1&0&2\\
	\end{bmatrix};  \mathcal{L}_2=\begin{bmatrix}
	3&-3&0\\
	0 &2&-2\\
	-1&0&1\\
	\end{bmatrix}
\end{align*}
Each cluster has only one agent, who is able to interact with agents outside its own cluster (agent 1 in the first cluster and agent 5 in the second cluster). The weights of the inter- cluster interactions are chosen as follows
\begin{align*}
P_e=\begin{bmatrix}
0.5&0&0&0&0&0.5&0\\
0&1&0&0&0&0&0\\
0&0&1&0&0&0&0\\
0&0&0&1&0&0&0\\
0&0.1&0.1&0&0.8&0&0\\
0&0&0&0&0&1&0\\
0&0&0&0&0&0&1\\
\end{bmatrix}
\end{align*}
According to \eqref{5}, the reset dynamics of the leader 1 (agent 1) and leader 2 (agent 5) are
\begin{align*}
&x_{l_1}(t_k^+)=0.5 x_{l_1}(t_k) +0.5 x_6(t_k), \;x_{l_1}=x_1\\
&x_{l_2}(t_k^+)=0.1 x_2(t_k)+0.1 x_3(t_k)+0.8 x_{l_2}(t_k), \;x_{l_2}=x_5
\end{align*}
and by by supposing that $x(0)=[0 -1 -2 -4 \;\; 2\;\; 3\;\; 4]^\mathsf T$. As a result, Fig. 1 depicts the state trajectories of agents in the clustered network and indicates that the MASs can reach consensus, which is consistent with the result of \textit{Theorem 1}.
\begin{figure}[htb]
	\centering
	\includegraphics[height=4.7cm]{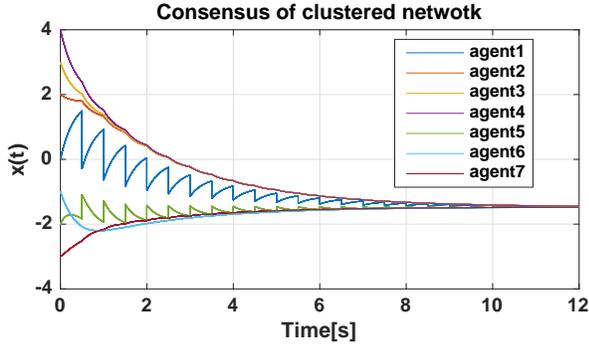}
	\caption{Consensus of 7 agents grouped in 2 clusters}
\end{figure}

Moreover, if the inter-cluster interaction is done by only leaders, who  interact at some discrete-time $t_k$ through predefined graph $\mathcal{G}_l=(\mathcal{I}, \mathcal{E}_l)$ corresponding to the matrix $P_l$
\begin{align*}
P_l=\begin{bmatrix}
	0.45 &0.55\\
	0.55 &0.45\\
\end{bmatrix} \Rightarrow w^T =[0.5\;\;0.5]
\end{align*}
and the eigenvectors of matrix $\mathcal{L}_1$ and $\mathcal{L}_2$ are
\begin{align*}
\mathsf{r}_1^T=[1/3\;\;1/3\;\;1/6\;\; 1/6];\mathsf{r}_2^T=[2/11\;\;3/11\;\;6/11]
\end{align*}
then by using \eqref{23} we obtain the following consensus value $x^{\star}=2.513$. Finally, we investigate the case consensus problem with external disturbance $w(t)=0.1*sin(2t)$. By choosing $\alpha =1, \rho=1$ and $\beta =0.7$, and solving LMI \eqref{31t} and \eqref{32} in \textit{Theorem 2}, one has
\begin{align*}
\small
\mathsf P=\begin{bmatrix}
2.36  &-0.69 &-0.42 &-1.25 &0&0&0\\
-0.69   &2.36 &-1.25  &-0.42&0&0&0\\
-0.42  &-1.25  & 1.97   &-0.30 &0&0&0\\
-1.25  &-0.42  &-0.30  & 1.96 &0&0&0\\
0&0&0&0&2.02   &-1.6 &-0.39\\
0&0&0&0&-1.6&    2.94& -1.30\\
0&0&0&0&-0.39  & -1.3  &1.69\\
\end{bmatrix}
\end{align*}
\begin{figure}[htb!]
	\centering
	\includegraphics[height=6.4cm]{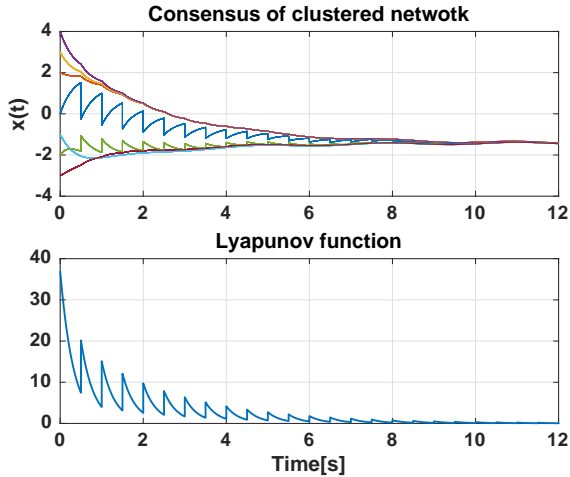}
	\caption{Consensus of 7 agents grouped in 2 clusters under disturbances}
\end{figure}

The result simulation is shown in the Fig. 2 and the evolution of the Lyapunov function $V$ is also depicted  in the Fig.2, which is consistent with the result of \textit{Theorem 2}.
\section{Conclusion}
In this paper, a novel approach has been proposed to design distributed consensus controllers for general linear MASs with the following features. First, the considered networks are partitioned into clusters, where the communication between agents inside each cluster is continuous, but the cluster's leader interacts outside its cluster at some reset times. Second, thanks to results from matrix theory and algebraic graph theory, the consensus problem in the clustered network is analyzed. Third, sufficient conditions for the robust $H_{\infty}$ stability of this equivalent system were derived from solutions of local convex LMIs problems, which can be solved in a distributed manner. A possible example of our proposed approaches was illustrated.
\bibliographystyle{IEEEtran}
\bibliography{refs}
                                                        
\end{document}